%% file: adbin.tex
\documentstyle[times,url,graphicx]{mn}

\include{defn}

\begin{document}

\title{Adaptive binning of X-ray galaxy cluster images}
\author[J.S. Sanders and A.C. Fabian] {J.S. Sanders and A.C. Fabian\\
  Institute of Astronomy, Madingley Road, Cambridge. CB3 0HA}
\maketitle

\begin{abstract}
  We present a simple method for adaptively binning the pixels in an
  image. The algorithm groups pixels into bins of size such that the
  fractional error on the photon count in a bin is less than or equal
  to a threshold value, and the size of the bin is as small as
  possible. The process is particularly useful for generating surface
  brightness and colour maps, with clearly defined error maps, from
  images with a large dynamic range of counts, for example X-ray
  images of galaxy clusters.  We demonstrate the method in application
  to data from \emph{Chandra} ACIS-S and ACIS-I observations of the
  Perseus cluster of galaxies. We use the algorithm to create
  intensity maps, and colour images which show the relative X-ray
  intensities in different bands. The colour maps can later be
  converted, through spectral models, into maps of physical
  parameters, such as temperature, column density, etc. The adaptive
  binning algorithm is applicable to a wide range of data, from
  observations or numerical simulations, and is not limited to
  two-dimensional data.
\end{abstract}


\section{Introduction}
A problem often encountered with the processing of images, especially
those from X-ray telescopes, is that there is a large dynamic range
present in intensity. In order to examine the structure of emission in
regions of an image where there are few photons, it is necessary to
bin the data using bins of large angular size. An alternative is to
smooth the data with a Gaussian kernel of large angular size.
Conversely, to examine structure in areas of high emission, a small
bin-size or kernel is sufficient. If one wishes to examine regions
with a wide variation in count-rate, then some procedure which has a
variable bin or kernel is required to create statistically optimal
maps.

Another related problem, which motivated the present work, is the
creation of optimal colour maps.  The ratio of photon counts in a low
intensity region will have a large associated statistical error, so
the counts need to be averaged over a large region to form the ratio.
The algorithm needs to account for the statistical accuracy in the
separate bands, not just the total image.

One technique which allows structure on a wide range of scales to be
revealed is adaptive kernel smoothing (Ebeling, White \& Rangarajan
2000; Huang \& Sarazin 1996). The \textsc{asmooth} algorithm of
Ebeling et al. (2000) convolves an image with a Gaussian kernel. If
the signal within the kernel, applied to part of an image, is of a
chosen minimum significance above the local background, then the
convolved signal is added to the output image. The kernel is increased
in size until all the counts in the input image are added to the
output image.

\textsc{asmooth} is very good at finding low surface brightness
features, such as filaments. It also especially useful for producing
images for display, and for identifying bright features against a
background.  It is not a universal smoothing tool, however, and care
needs to be taken before using it in a particular situation. For
instance, it is designed to identify positive features against a
background, but not to identify `holes' in emission. Due to the way it
identifies significance against the local background, it may produce
spurious features in areas where there is bright flat emission, for
instance the cores of galaxy clusters, where Poisson noise may become
significant.  Used with care, \textsc{asmooth} is a useful routine in
many situations.

The analysis method outlined in this paper, \emph{adaptive binning},
was originally developed to produce temperature, column density and
colour maps of the Perseus cluster in Fabian et al. (2000). We
required a simple algorithm to compare the relative fluxes in
different X-ray bands to a theoretical plasma emission model. In order
to be able to spatially resolve the ratios, it is necessary to bin or
smooth the data.  Binning, rather than smoothing, is a better method
for producing spatial estimates of counts for analysis, because it
does not spread counts around an image beyond the boundaries of the
bin.

In this paper we use data analysed from the 24\,ks ACIS-S
\emph{Chandra} X-ray observation of the Perseus Cluster, Abell\,426,
from Fabian et al. (2000), to demonstrate the adaptive binning method.
The Perseus Cluster is the brightest X-ray cluster in the sky, and
contains a large cooling flow (Fabian et al. 1994) with a mass
deposition rate of approximately $300\Msunpyr$. Interesting
substructure in the cluster due to its central radio source makes it
an ideal object for our purpose.  We will not attempt here to analyse
the physics behind the observations, which is left to Fabian et al.
(2000) and further work. We also study the method by applying it to
simulated data.

The algorithm is presented in a logically consistent way, from
intensity adaptive binning to colour adaptive binning. Colour binning,
as mentioned above, was our original motivation, developed to analyse
X-ray colour cluster images.

\section{Intensity binning}
Adaptive intensity binning is the simplest case of the algorithm. It
attempts to adaptively bin a single image based on the number of
photons in each region.  The basic method is to bin pixels in
two-dimensions by a factor of two, until the fractional Poisson error
of the count in each bin becomes less than or equal to a threshold
value. When the error is below this value, those pixels are not binned
any further. The algorithm, in detail, is as follows.
\begin{enumerate}
\item Each pixel in the image is put into its own `bin', the term we
  use for a collection of pixels. A pixel here means one of the
  individual picture elements which form the input image. Essentially
  the image is initially divided into imaginary $1\times 1$ pixel bins.
\item If there are $n_i$ pixels in bin $i$, the total count in the bin
  is $c_i$, and the background per pixel is $b$, the net count in the
  bin is simply defined by
  \begin{equation}
    s_i \equiv c_i - n_i b.
  \end{equation}
  \label{item:start}
\item The fractional error on the net count in the bin is
  \begin{equation}
    \frac{{\sigma}_{\left( s_i \right) }}{s_i} =
    \frac{\sqrt{c_i + n_i b}}{c_i - n_i b}.
    \label{eqn:counterror}
  \end{equation}
  This is also the error on the average count in each pixel.
\item If the fractional error is less than or equal to a threshold
  value, then the pixels in the output image which correspond to the
  pixels in the input bin are set to the average mean count, $s_i /
  n$. The fractional error of the net count in the bin is also stored
  in the pixels in an `error image'. Additionally the bin is marked as
  having been processed.
\item Each bin is merged into its neighbouring three bins, to make new
  bins containing $2 \times 2$ of the previous bins. The four bins
  with the lowest $x$ and $y$ coordinates (lowest declination and
  highest right-ascension) are merged, as is each consecutive set of
  four bins. Any bins which have already been processed are ignored in
  the merging. It is useful to remember we are considering a bin as a
  list of pixels. Pixels which have already been set in the output
  image are ignored in future iterations.
\item The process is repeated from \ref{item:start} until there is
  only a single bin remaining.
\item A `bin-map' is also produced by the algorithm, giving an
  identification number for each processed bin in terms of the pixels
  which it contains. Using the bin-map, any image of that size can be
  binned using the same bins.
\end{enumerate}

\subsection{Demonstration}
\begin{figure}
  \begin{center}
    \includegraphics[width=0.8\columnwidth]{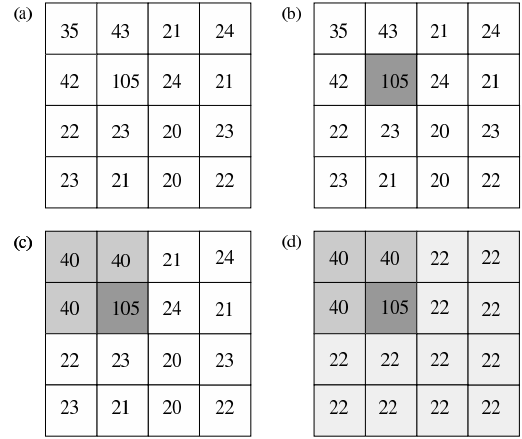}
  \end{center}
  \caption{Example demonstrating the intensity adaptive binning
    process. The bottom-left pixel is the origin, $(0,0)$. See the
    text for details.}
  \label{fig:bin_dia}
\end{figure}

A simple demonstration of the algorithm operating on a $4 \times 4$
pixel image is shown in Fig. \ref{fig:bin_dia}. We demonstrate
the process using a threshold fractional error of 0.1 and no background
counts. (a) shows the image before binning. We first look for
individual pixels (bins of $1\times 1$ pixels) with a Gaussian
fractional error in the count less than or equal to the
threshold. Only one pixel does; it is shown `painted' in (b). We then
bin the remaining pixels by a factor or 2. We examine the binned pixels
to see whether any have errors less than the threshold. The three
pixels in the top-left corner do, so they are averaged together and
painted in (c). We then bin again with a $4 \times 4$ pixel bin. The
remaining pixels have an error less than the threshold, so they are
averaged and painted in (d), the final output image. Had the error on
the final bin been larger than the threshold, it would have
been binned anyway.

\subsection{Real example}
\begin{figure}
  \includegraphics[width=0.95\columnwidth]{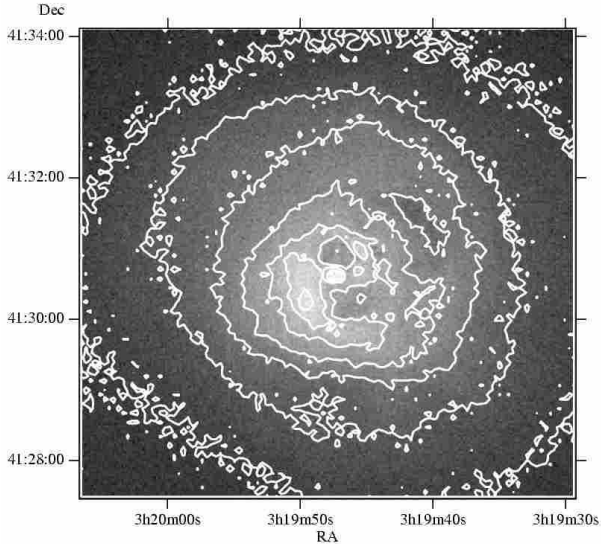}
  \caption{Image of cluster binned with 2 arcsec pixels. The image has
    been corrected for exposure and readout node lines. 8 contours
    are spaced on a square-root scale, between 20 and 300 counts per
    pixel.}
  \label{fig:per_raw}
\end{figure}

We show in Fig. \ref{fig:per_raw} a `raw' image of the Perseus cluster
in the 0.5--7\,keV band from the \emph{Chandra} observation, binned
with 2 arcsecond pixels. The image has been exposure-corrected. It
also has been corrected for the readout node lines of the ACIS-S
detector (this smears out the point source at the
centre). Contours were placed on the image showing 8 levels spaced
equally in square-root space between 20 and 300 counts.

The Perseus Cluster is very bright, so we see much structure in the
data without doing more than simple binning.  The contours shown have
been smoothed, but even they break up at large radii due to the count
rate being swamped by Poisson noise.

\begin{figure}
  \includegraphics[width=0.95\columnwidth]{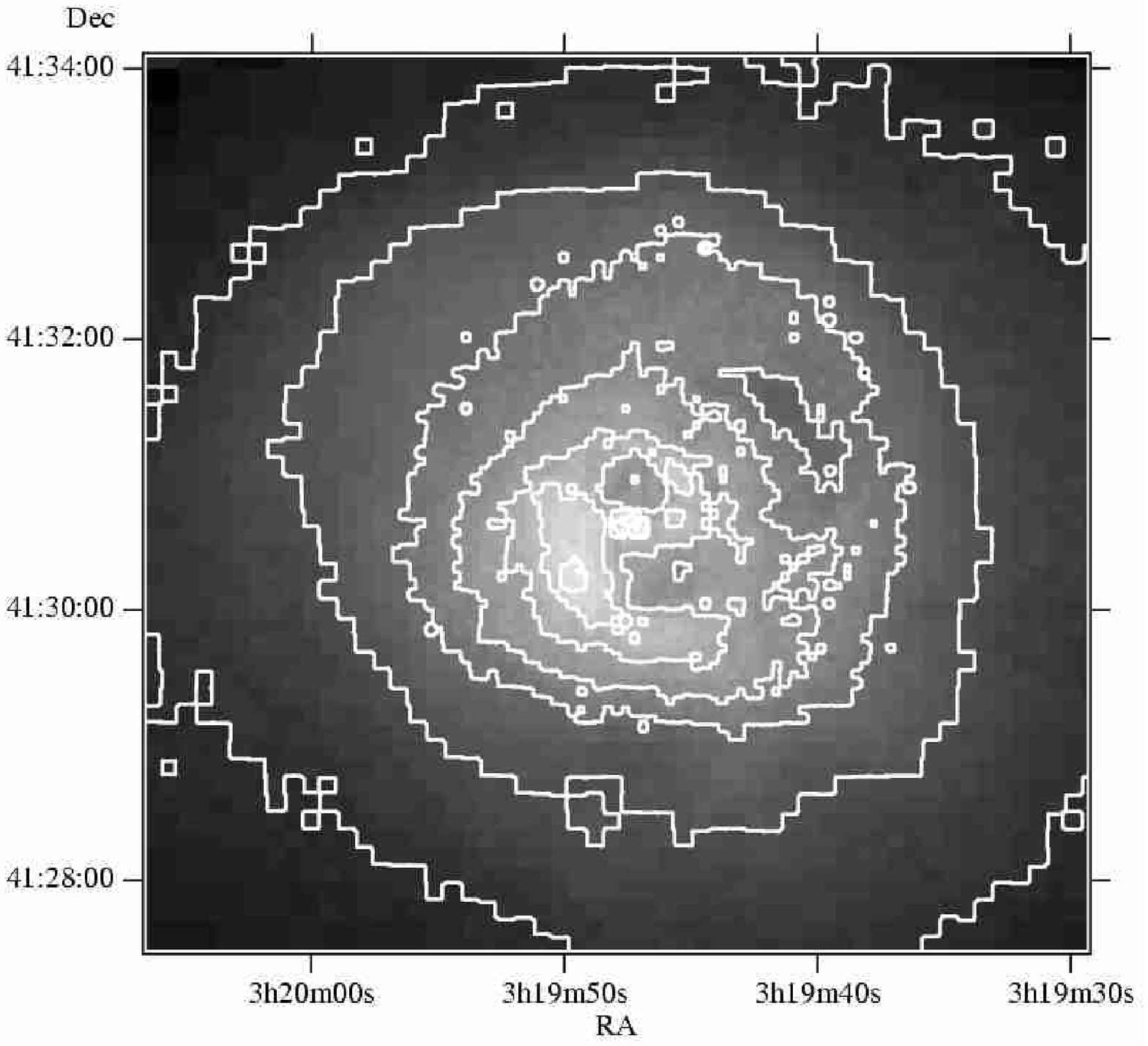}
  \caption{Adaptively binned image of the cluster. The pixel fractional
    error is set as 0.06. The contours are at the same levels as
    Fig. \ref{fig:per_raw}.}
  \label{fig:per_adbin6}
\end{figure}

Fig. \ref{fig:per_adbin6} shows the raw image of Perseus adaptively
binned with a threshold fractional error of 0.06.  The X-ray
background was ignored since it was low, even near the image edges.
The contours on the image near the centre match the contours in the
raw data well, showing the algorithm works in this regime. At the
sides, the bins are quite large and blocky, but otherwise there do not
appear to be any edge-effects. There is still noise present in the
image, but at a level much reduced from the raw data.

\begin{figure}
  \includegraphics[width=0.95\columnwidth]{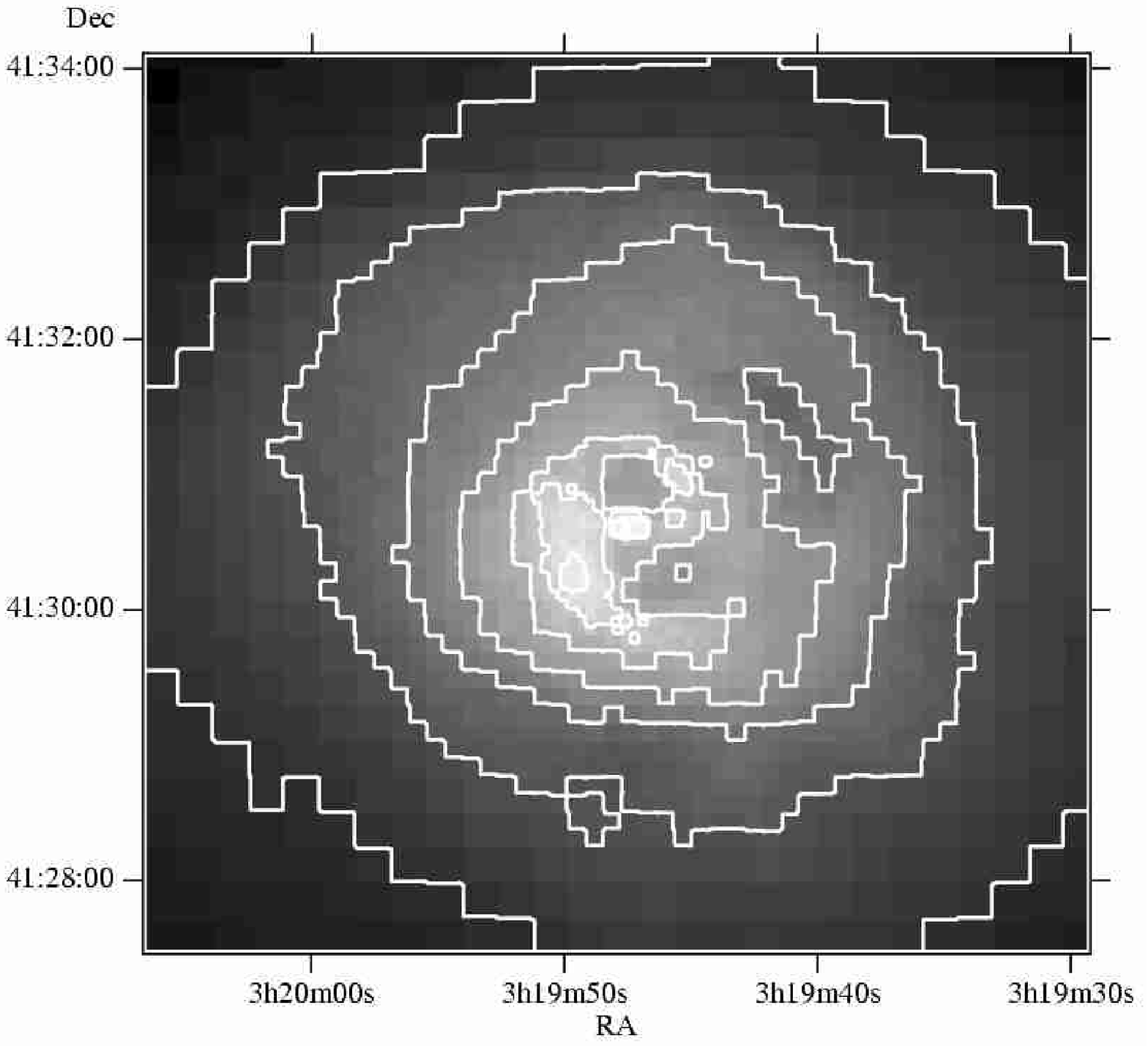}
  \caption{Adaptively binned image of the cluster. The pixel fractional
    error is set as 0.04. The contours are at the same levels as
    Fig. \ref{fig:per_raw}.}
  \label{fig:per_adbin4}
\end{figure}

Fig. \ref{fig:per_adbin4} shows the cluster again adaptively binned,
but now with a threshold error of 0.04. Note how the bins are larger,
but the level of noise is significantly less. A real feature has been
lost from the image, however. A point source present in the upper left
radio lobe has disappeared. Its counts were merged into rest of the
emission from that region.

\begin{figure}
  \includegraphics[width=0.95\columnwidth]{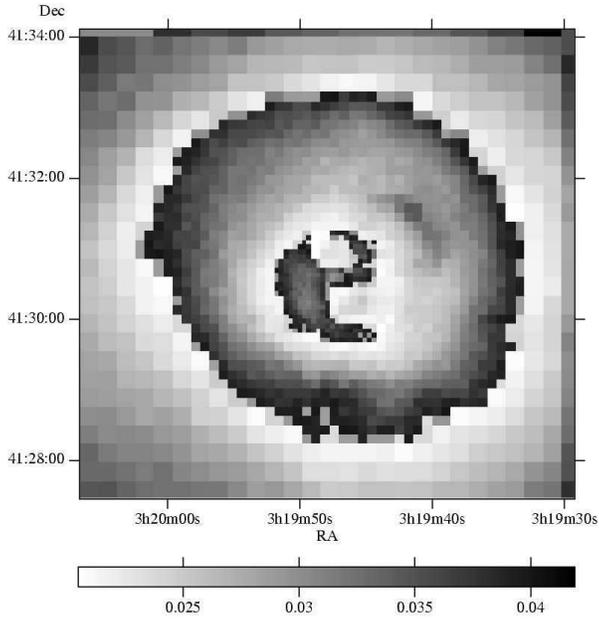}
  \caption{Fractional error map for adaptively binned image in
    Fig. \ref{fig:per_adbin4}.}
  \label{fig:per_adbin4_err}
\end{figure}

In Fig. \ref{fig:per_adbin4_err} is shown the error map for the
adaptively binned map above. In it are easily visible the changes in
bin size as the count rate decreases towards the outside. As the count
rate decreases, the error of the bins increases until it reaches the
threshold, and then the bin-size doubles.

\begin{figure}
  \includegraphics[width=0.95\columnwidth]{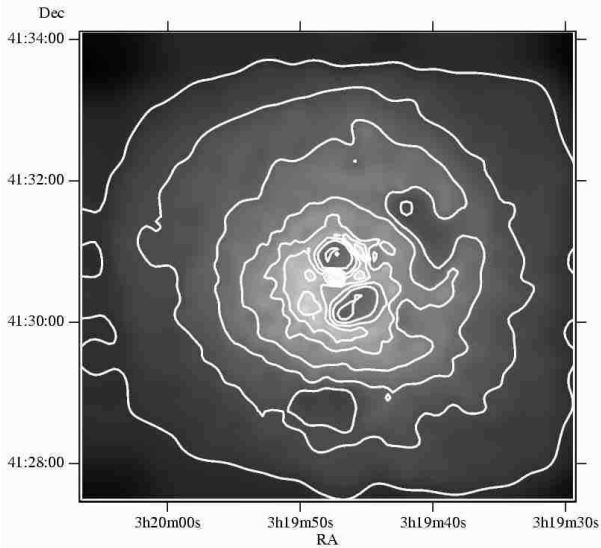}
  \caption{Adaptively smoothed image of the cluster, with a minimum
    significance of 4--$\sigma$. The contours are at the same levels as
    Fig. \ref{fig:per_raw}. Note the edge effects in the contours.}
  \label{fig:per_asmooth}
\end{figure}

For comparison, Fig. \ref{fig:per_asmooth} shows an adaptively
smoothed (AS) image of the cluster calculated from the data in
Fig. \ref{fig:per_raw}. It was smoothed with a minimum significance of
4--$\sigma$ using the \textsc{asmooth} algorithm of Ebeling et al. The
contours are at the same levels as the raw image.

The AS image contains sharp positive features, such as the edges of
the radio lobes. It does not perform as well in the determination of
the number of counts in the negative features, such as the radio lobes
themselves, where the number of counts per pixel is half that of the
raw data. However, the algorithm is designed to find positive
features, so using it to look at holes is a misapplication.  Also
apparent are some edge effects, where \textsc{asmooth} does not find
enough counts to place a high significance on the generated
features. The edge effects are avoidable by smoothing a larger
area of sky than required, and cropping the image thereafter. One
disadvantage of this is that \textsc{asmooth} running time increases
quickly with image size.

\begin{figure}
  \includegraphics[width=0.95\columnwidth]{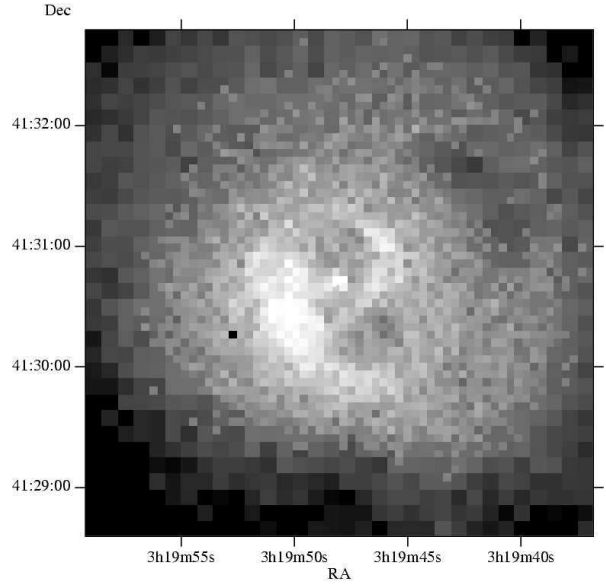}
  \caption{ACIS-I adaptively binned image of the Perseus cluster,
    using a fractional error of 0.15.}
  \label{fig:per_acisi_ab}
\end{figure}

In Fig. \ref{fig:per_acisi_ab} we show another adaptively binned image
of the Perseus cluster. This, however, was created from data from an
observation using the ACIS-I detector on \emph{Chandra}, with an
exposure of 18.6\,ks. We present it because the number of counts per
pixel is lower than the ACIS-S image, and so it is a good
demonstration of the algorithm used on data with more noise.

Adaptive binning of the ACIS-I image brings out an interesting feature
of the cluster; it clearly demonstrates that the southern rim of the
northern radio lobe (hole) lies south of the nucleus. There is a linear
diagonal structure present to the south of the nucleus, running from
the north-west to the south-east. This structure is also present in
the raw data, but is not clear on the ACIS-S image, despite the
longer exposure, as the ACIS-S raw image contains a dark strip due to
the node-line of the detector, where the effective exposure
is short.

The dark bin to the south-east of the nucleus shows a `stranded bin',
an occasional problem with the algorithm, which we discuss later.

\subsection{Notes on the algorithm}
\subsubsection{General}

\begin{enumerate}
\item The algorithm allows bins of width $n$ to contain pixels with
  coordinates $(x,y)$ in the range $ni \le x < n(i+1)$, $nj \le y <
  n(j+1)$, where $i$ and $j$ are integers, and $i \ge 0$, $j \ge 0$.
  The coordinate origin is $(0,0)$. $x$ and $y$ increase in the
  directions of decreasing right-ascension and increasing declination,
  respectively.
  
\item If the unbinned image does not have sides which are an equal
  length of pixels, or are not a power of two, then the bins on one
  or both sides will be truncated. This can lead to bins with a small
  numbers of counts and a large error on the final pass.
  
\item As is the case with conventional binning, sources with small
  spatial extent may be split between two bins if they cross a bin
  boundary. This can be solved by allowing sub-bin positioning,
  described as follows. For each pass, bins are allowed to be placed
  at positions displaced by a multiple of some fraction of their
  length from their conventional positions. The errors are calculated
  for each possible bin, and the bins are sorted into order of
  ascending error, discarding those with an error larger than the
  threshold. The bins are painted in that order, ignoring those bins
  which had already been partially painted.
  
  The disadvantage of this technique is that the output image appears
  `dithered'. Bins are painted close to one another, but without
  enough space to put another bin between them. The result is a loose
  cluster of small bins inside a larger one. The original algorithm
  appears to produce better results and is less dependent on small
  variations of errors.
  
\item The bins are doubled in size between each pass, rather than
  slowly increased in size, as this ensures a whole number of smaller
  bins fit inside a larger bin. If this is not the case, then gaps are
  left between bins in the binning, leading to dithering.
  
\item Allowing rectangular bins is not useful. It is ambiguous in
  which direction a rectangle could lie, or be extended. Their use
  could introduce spurious linear structures into the output image.
  
\item The use of non-square bins is a possible technique. Any two
  dimensional object that tiles together could be used, but it is not
  clear what is the optimal shape. Square bins are a good choice for a
  number of reasons, including the fact that pixels are naturally
  square in most detectors; square bins act as scaled pixels.
  However, there is probably some room for additional work to find
  better shapes.  An alternative idea is to use contour levels in an
  AS image to define `bins', as demonstrated by Sanders, Fabian \&
  Allen (2000).
\end{enumerate}

\subsubsection{Overlayed colour images}
The adaptive-binning algorithm is useful for producing images
demonstrating in real colour where the soft, medium and hard areas of
emission are. First an intensity image can be binned to produce a
bin-map. Then images in three energy bands can be binned themselves
using the bin-map. A software package such as \textsc{gimp} can then
be used to combine the three images as, for example, the red, green
and blue layers in a single image.  The contrast of the image can then
be increased to highlight the areas of hard and soft emission.

The advantage of doing this rather than using raw data to make the
image, is that the low-intensity regions are not dominated by noise.
The technique was used to create the colour image of the Perseus
cluster in Fabian et al. (2000).

\subsubsection{Stranded pixels and contiguous regions}
One `feature' of the algorithm is that pixels may become `stranded'.
If a bin has a reduced count relative to its neighbours, it may be
left out on a pass and have no nearby pixels to be binned with on the
next.  It may be left until the final pass, and that area will be
binned with the remaining pixels. Such pixels could be cosmetically
removed by replacing them with their original value.

A modified version of the algorithm avoids stranded pixels. If a bin
consists of more than one non-contiguous sets of pixels, then the bin
is split into several contiguous regions, and each region treated by
itself. A stranded set of pixels will remain until the end, where it
will be binned by itself, as it is isolated.  Here, two pixels are
said to be connected if one pixels is one of the neighbouring eight
pixels of the other pixel.

This modification has a couple of disadvantages. Firstly, finding
contiguous sets of pixels is fairly slow. Also, parts of the image
which would be connected were the image larger, are treated by
themselves, and therefore have a larger statistical error on their
count. With these problems in mind, it is probably a useful
modification for many applications of the algorithm. For simplicity,
we will use the unmodified version of the algorithm in the following
sections.

\subsection{Simulated cluster}
\begin{figure}
  \begin{tabular}{llll}
    (a)&(b)\\
    \includegraphics[width=0.46\columnwidth]{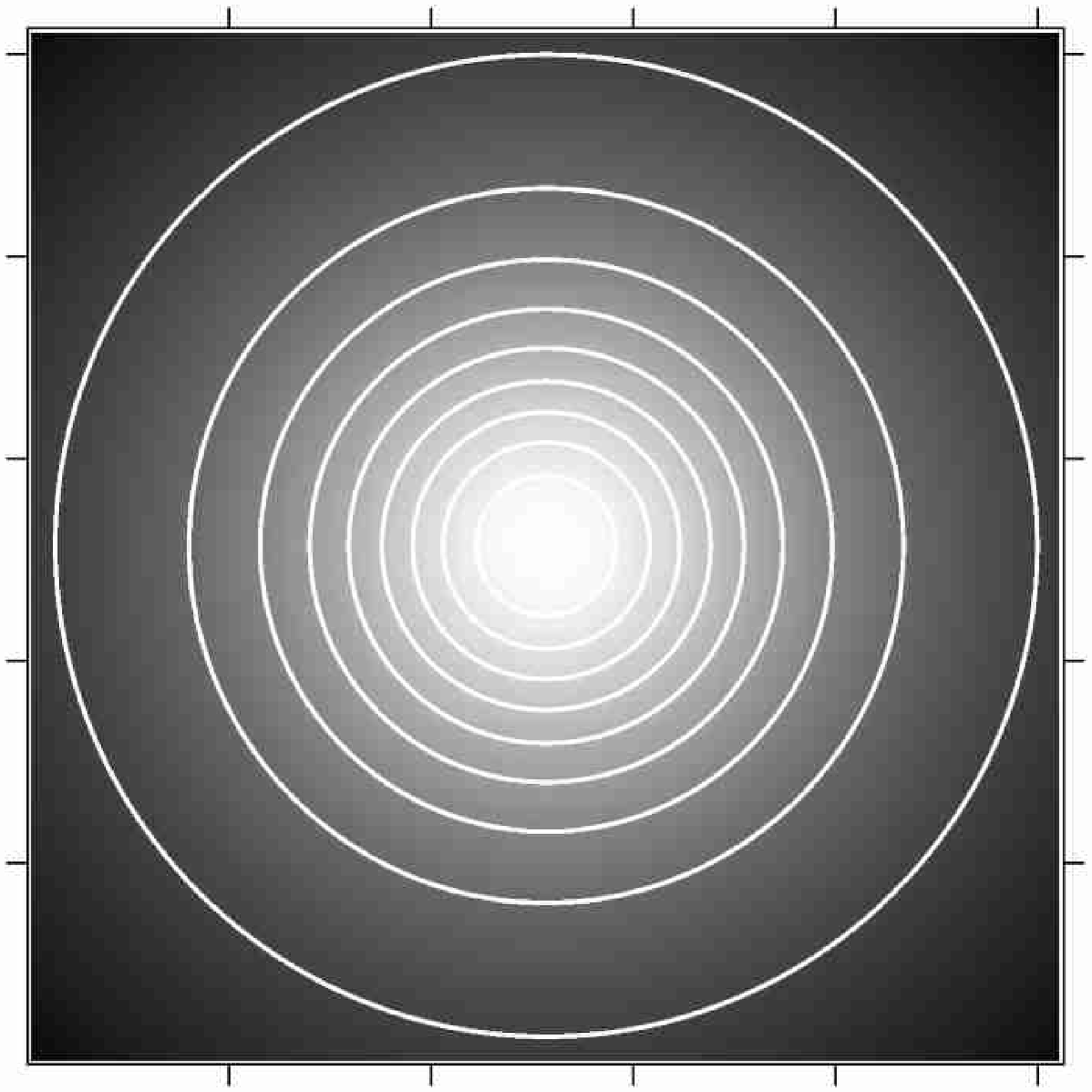} &
    \includegraphics[width=0.46\columnwidth]{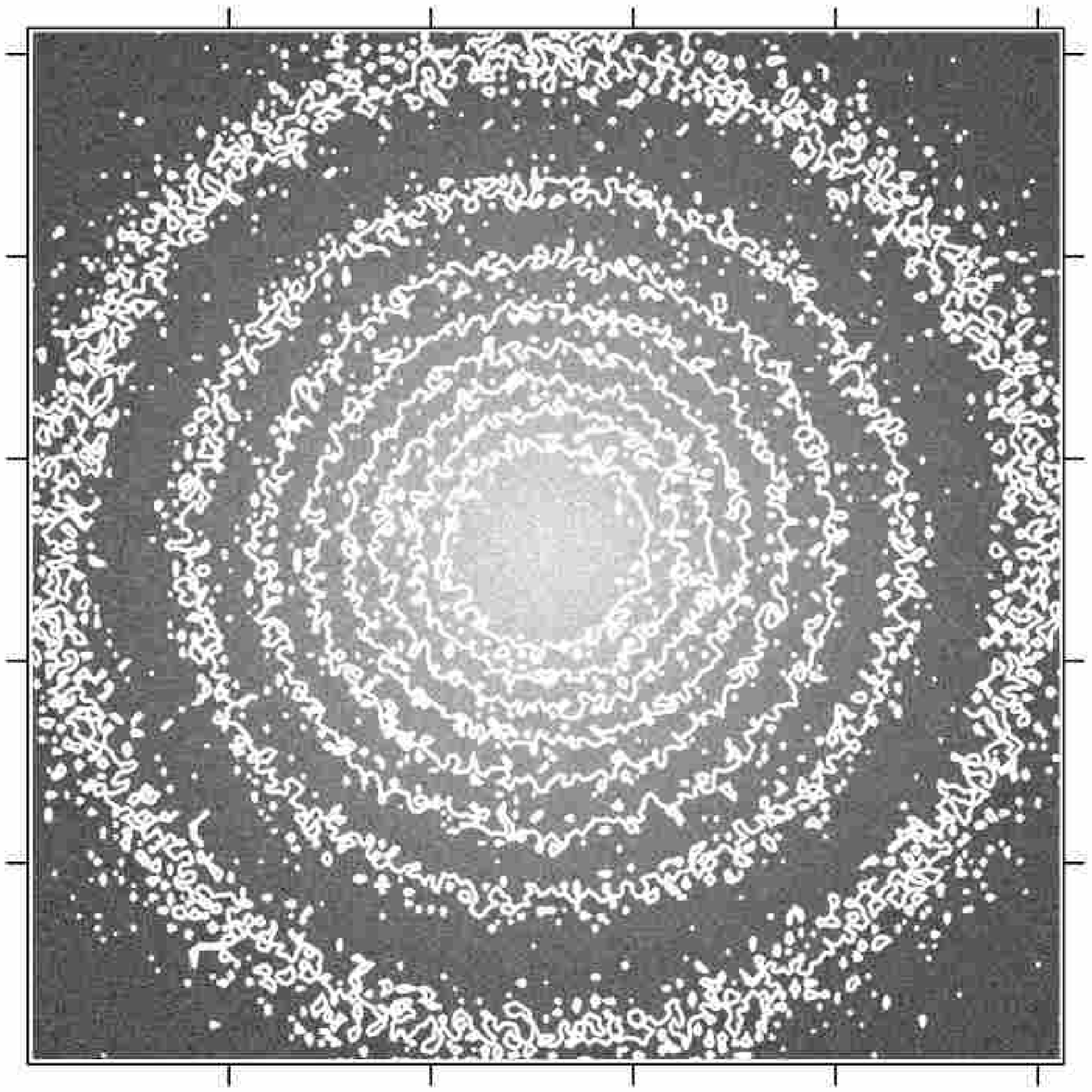} \\
    (c)&(d)\\
    \includegraphics[width=0.46\columnwidth]{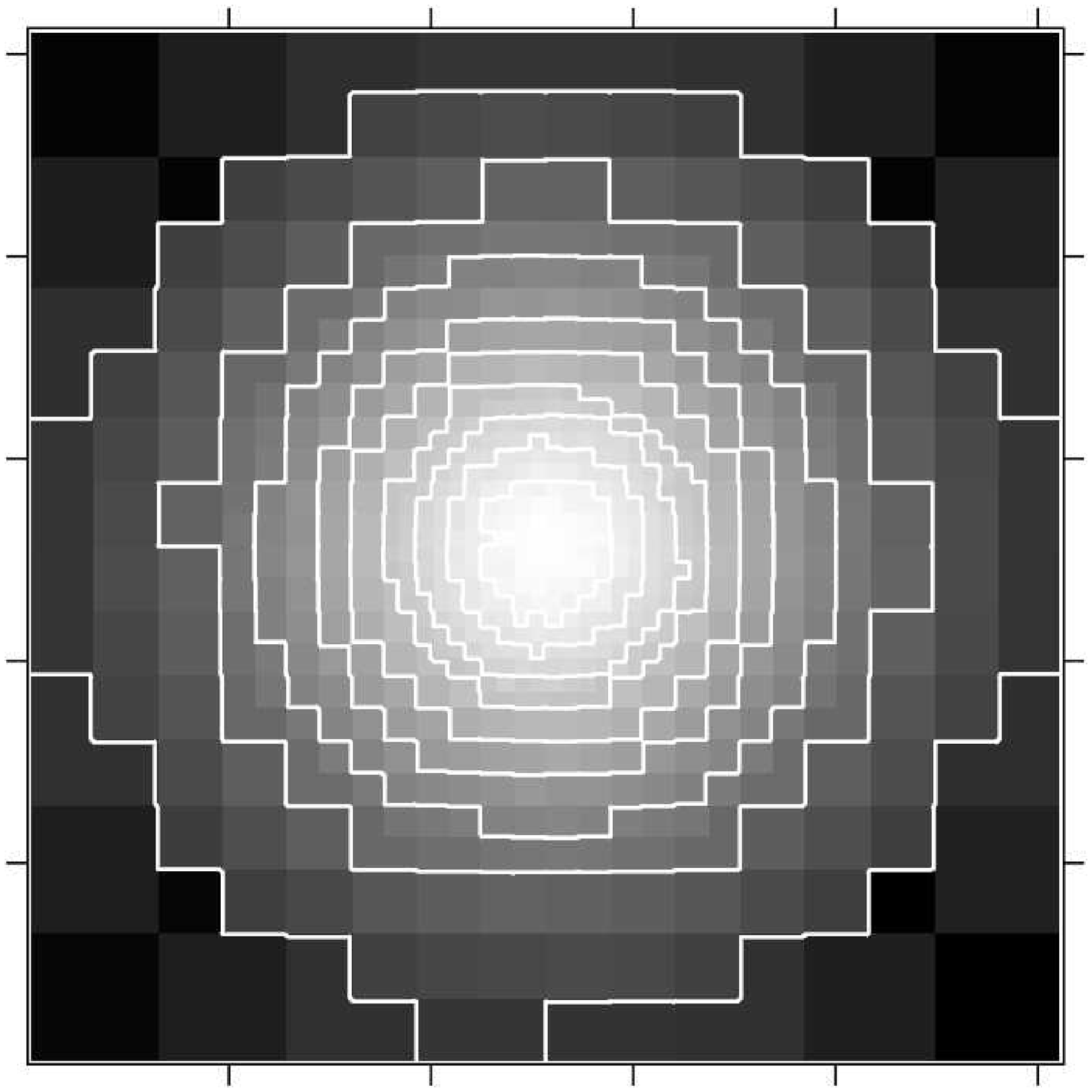} &
    \includegraphics[width=0.46\columnwidth]{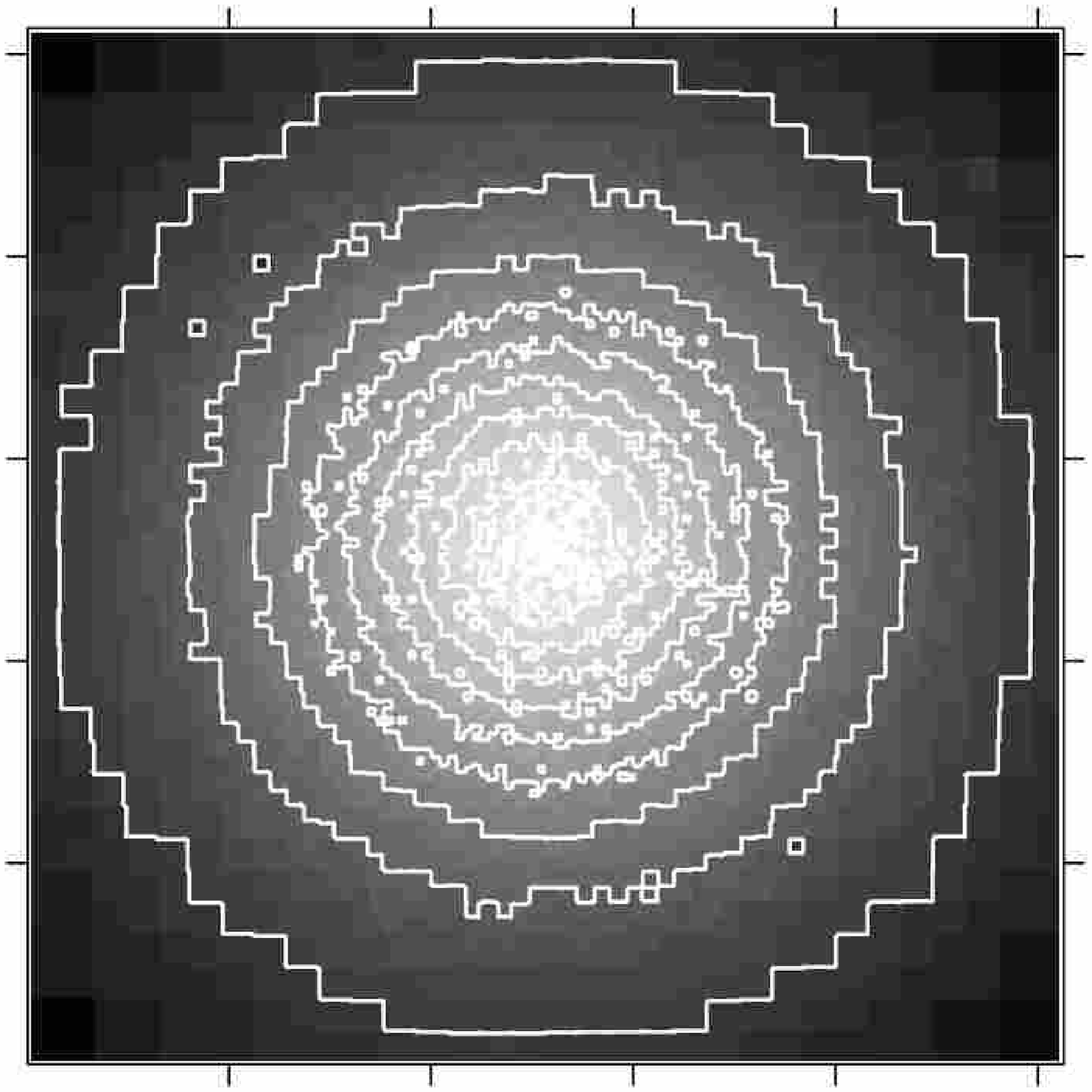} \\
    (e)\\
    \includegraphics[width=0.46\columnwidth]{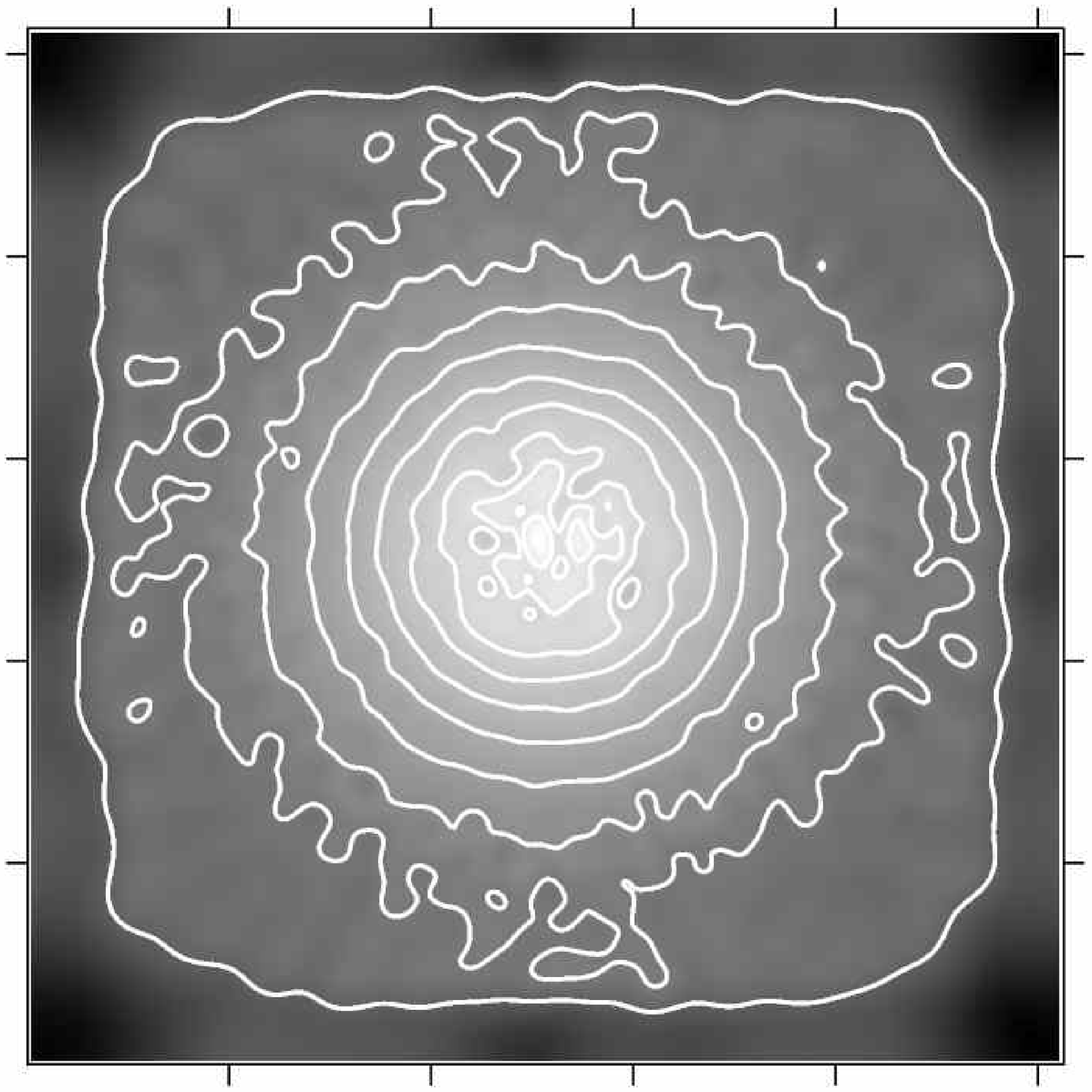}
  \end{tabular}

  \caption{Simulated cluster images. (a) Model surface brightness,
    contours range 10-90. (b) Simulated observation of cluster,
    contours range 30-100. (c) Adaptively binned image of cluster with
    error threshold of 0.02, contours range 10-90. (d)
    Adaptively binned image of cluster with error threshold of 0.06,
    contours range 10-100. (e) Adaptively smoothed image of
    cluster, $\sigma_{\mathrm{min}}=4$, contours range 10-100.}
  \label{fig:simcluster}
\end{figure}

In order to properly test the properties of the algorithm, we
attempted to simulate a simple image of a cluster. We used a
$\beta$-model to simulate the cluster with the following form (Sarazin
1988):
\begin{equation}
  S_b(r) = S_0 \left[ 1 + \left( \frac{r}{r_c} \right)^2 \right]^{0.5-3 \beta},
\end{equation}
where $S_b$ is the surface brightness at a radius $r$, $r_c$ is a
critical radius and $\beta$ is a parameter. We created an image of
$512 \times 512$ pixels, where $r_c$ was 128 pixels, $S_0$ was 100
counts/pixel, and $\beta$ was 0.67. The meaning of the word count, as
used here for surface brightness, is the expectation value of an
observation, and is allowed fractional values. The cluster was
positioned in the centre of the image. It is shown in Fig.
\ref{fig:simcluster}(a). The contours show constant surface brightness
from 10-90 counts in linear 10 count intervals.

To simulate how such a cluster would appear, if observed, we took the
surface brightness at a particular pixel, and randomly generated a
value from a Poisson distribution with that surface brightness as the
expectation value, to which we added a background with an expectation
value of 20 counts/pixel. The Poisson image is shown in Fig.
\ref{fig:simcluster}(b), with contours between 30 and 100 with the 10
count intervals. The contours are further out because of the
background. The data was smoothed in contouring, as it was too noisy
otherwise.

We processed the Poisson simulated data with our adaptive binning
algorithm to produce Fig. \ref{fig:simcluster}(c). Here a fractional
error threshold of 0.02 was used.  Fig. \ref{fig:simcluster}(d) shows
the output of the method using a threshold of 0.04, i.e. twice as
large a threshold as (c).  The contours are at the same levels as the
original surface brightness image.

For comparison, we show the output from the \textsc{asmooth} program
using a minimum feature significance of 4-$\sigma$ in Fig.
\ref{fig:simcluster}(e). \textsc{asmooth}, by default calculates the
local background and does not take a global value. To create this
image we smoothed the simulated data and subtracted the background
manually. The count per pixel becomes negative in the outer regions,
as \textsc{asmooth} does not know about the global background, which
also means features shown may have a lower significance than
4-$\sigma$.

\subsubsection{Spatial distributions of differences}
\begin{figure}
  \begin{tabular}{llll}
    (a)&(b)\\
    \includegraphics[width=0.46\columnwidth]{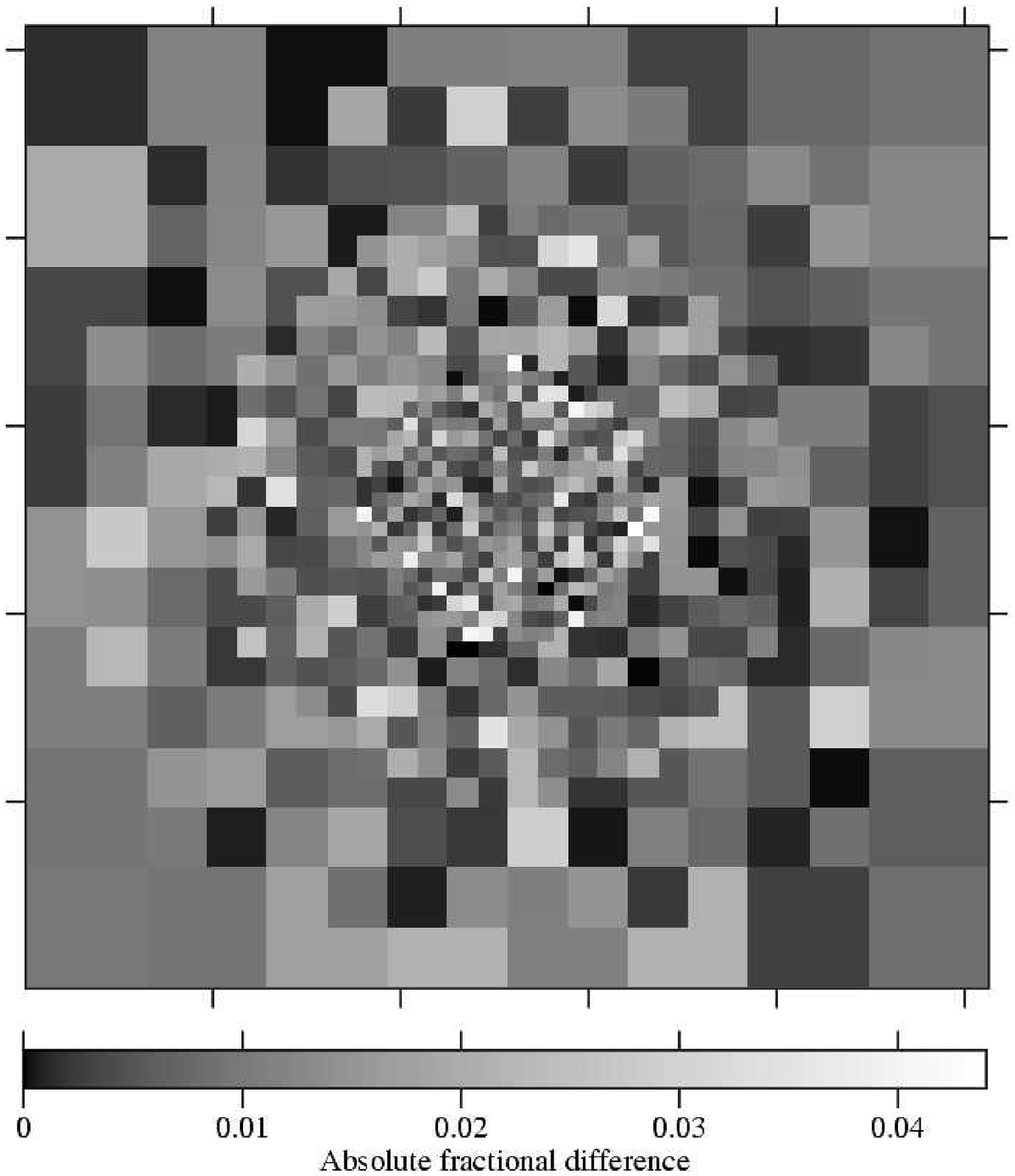} &
    \includegraphics[width=0.46\columnwidth]{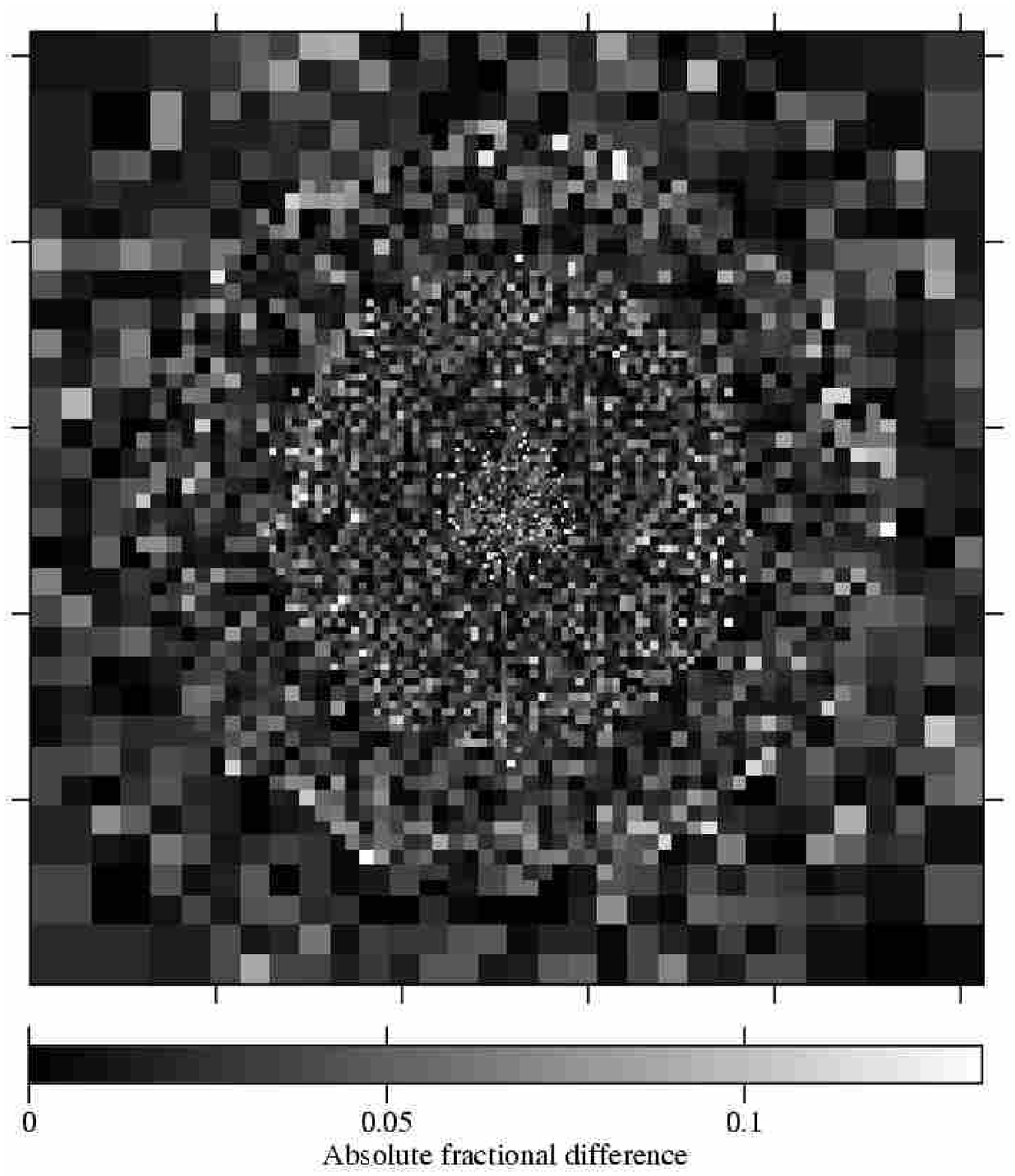} \\
    (c)&(d)\\
    \includegraphics[width=0.46\columnwidth]{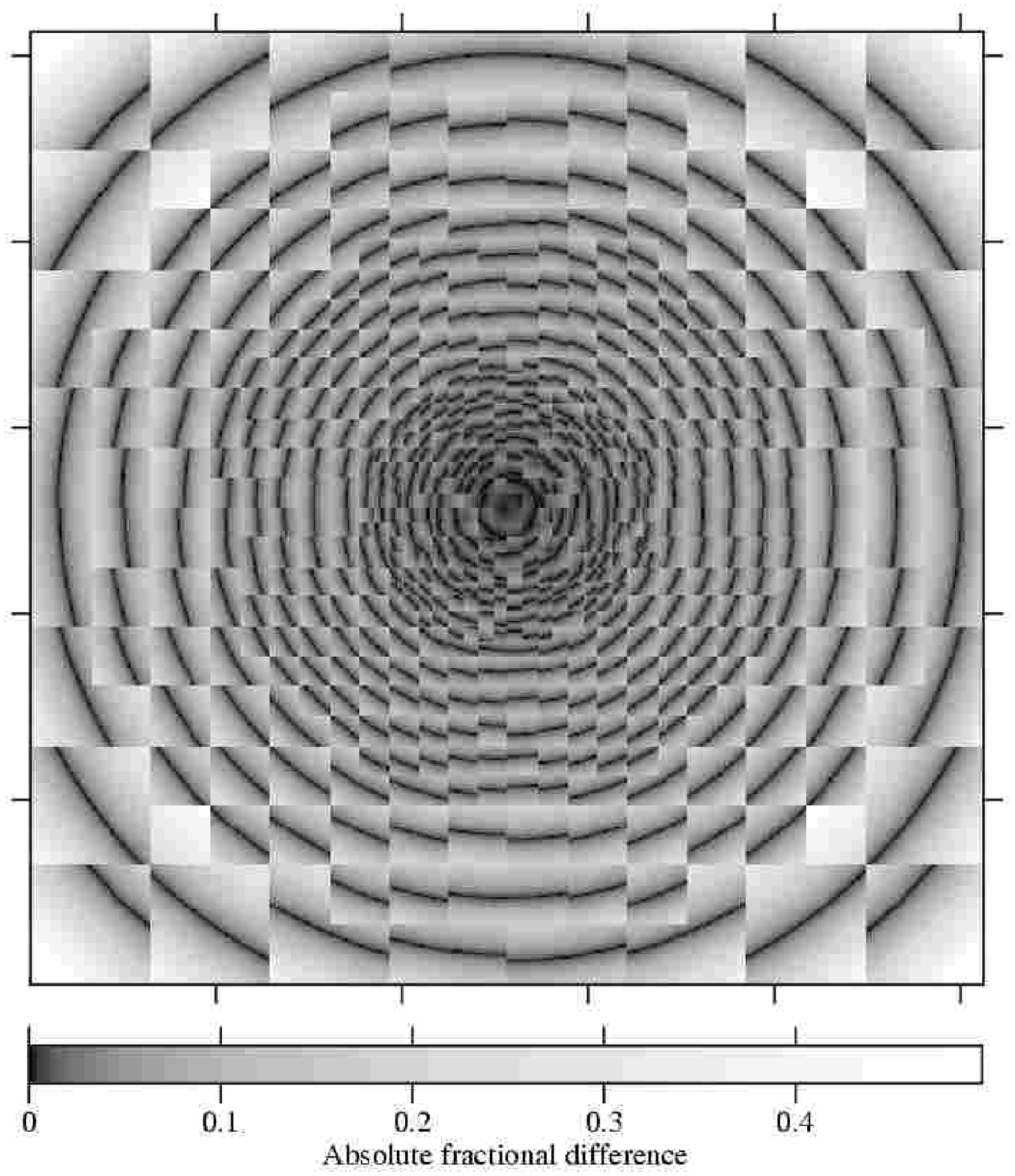} &
    \includegraphics[width=0.46\columnwidth]{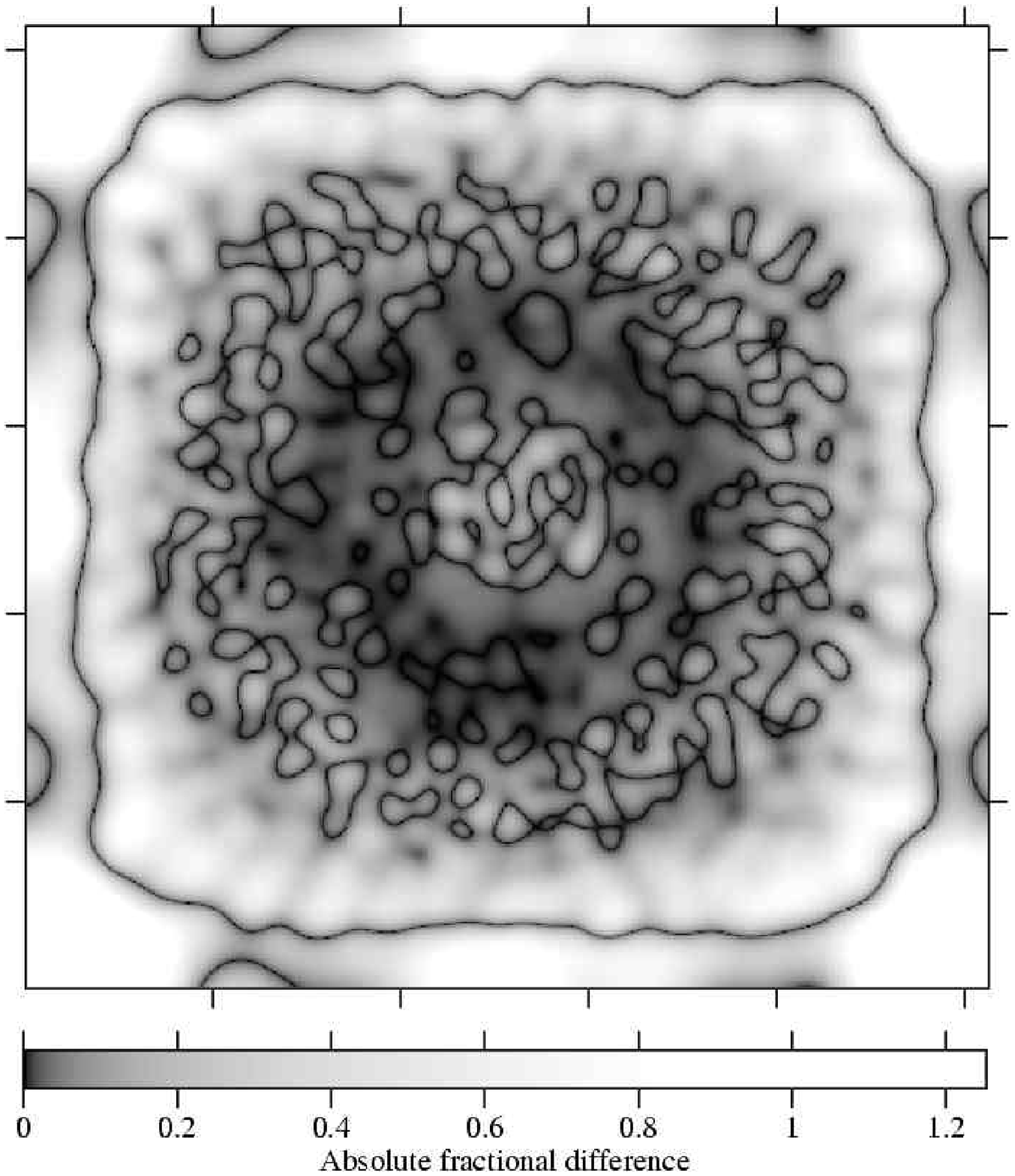} \\
  \end{tabular}

  \caption{Absolution fractional differences between the
    reconstructed images and original surface brightness image.  (a)
    Adaptively binned image, error 0.02, and binned surface brightness
    image. (b) Adaptively binned image, error 0.06, and binned surface
    brightness image.  (c) Adaptively binned image, error 0.02 and
    unbinned surface brightness image.  (d) Adaptively smoothed image,
    $\sigma_{\mathrm{min}}=4$ and unbinned surface brightness image.}
  \label{fig:simcluster_delt}
\end{figure}

We constructed in Fig. \ref{fig:simcluster_delt} images displaying the
absolute fractional differences between the reconstructed and model
surface brightness images. Spatial correlations in errors should
be visible on the images.

Fig. \ref{fig:simcluster_delt}(a) shows the absolute differences
between an adaptively binned image with 0.02 fractional errors (Fig.
\ref{fig:simcluster}(c)), and the original surface brightness image
binned with the same bins. (b) shows the same except using bins
constructed for 0.06 fractional errors (Fig. \ref{fig:simcluster}(d)).
There do not appear to be any correlations between the two images, in
terms of where the differences are large and small. The differences
appear noise-like and random. Note, however, that if the adaptively
binned image were binned with inappropriate bins, then this would not
show in (a) and (b) due to the surface brightness image being binned
with the same bins.

Fig. \ref{fig:simcluster_delt}(c) is an image showing the absolute
differences between adaptively binned image with 0.02 fractional
errors and the unbinned surface brightness image. Due to the count
rate varying across the bin, only the centres have a low absolute
fractional difference. This would be the same for any binning
process.

Fig. \ref{fig:simcluster_delt}(d) shows the differences between the AS
image in \ref{fig:simcluster}(e) and the surface brightness image for
comparison. Visible are the edge effects of the algorithm. Also
present are some peaks at the cluster core. These are due to the local
background estimation of the algorithm measuring significant
detections against the flat core of the cluster centre.

\subsubsection{Histograms of differences}
\begin{figure}
  \begin{tabular}{ll}
    (a)&(b)\\
    \includegraphics[width=0.46\columnwidth]{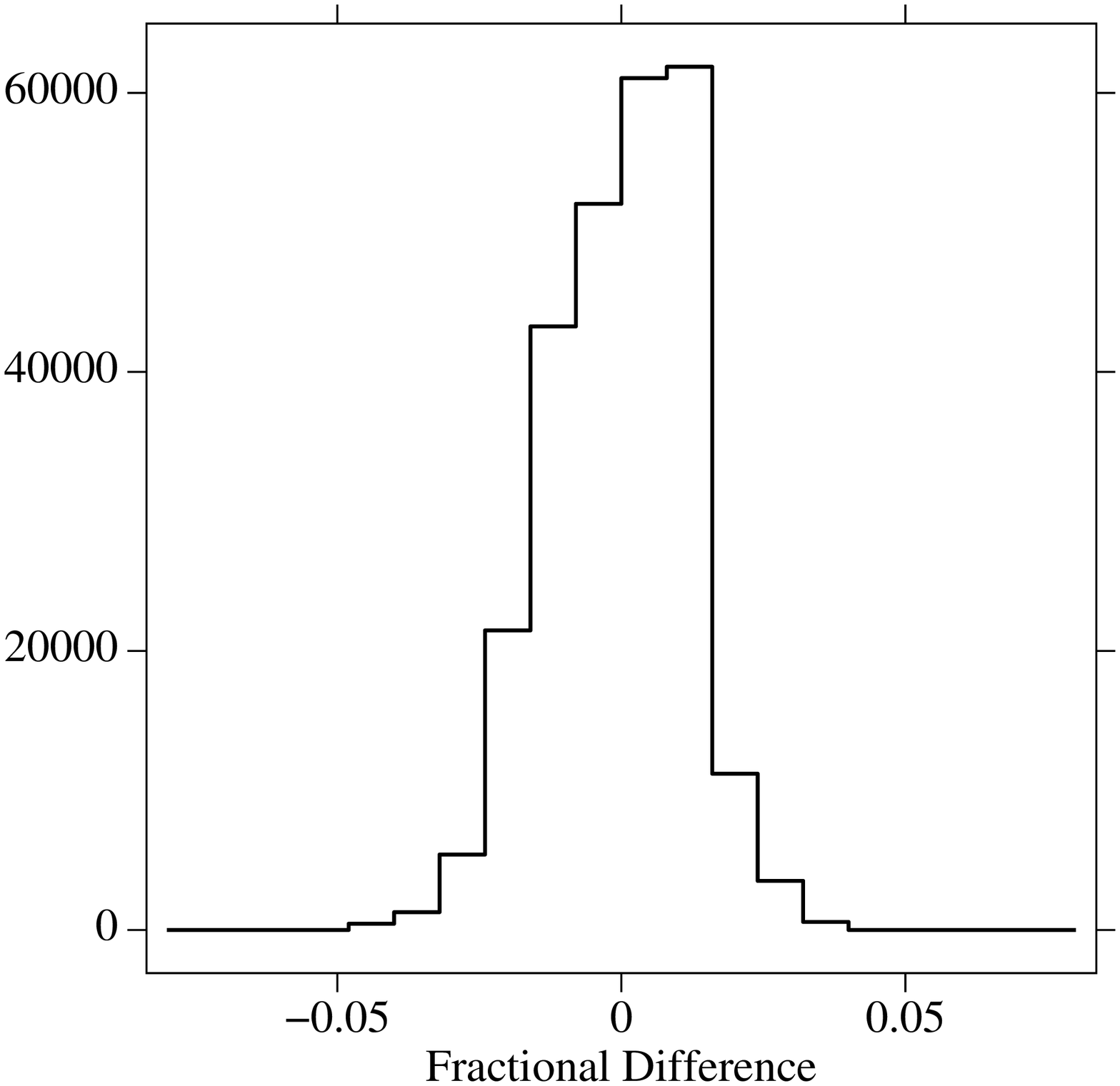} &
    \includegraphics[width=0.46\columnwidth]{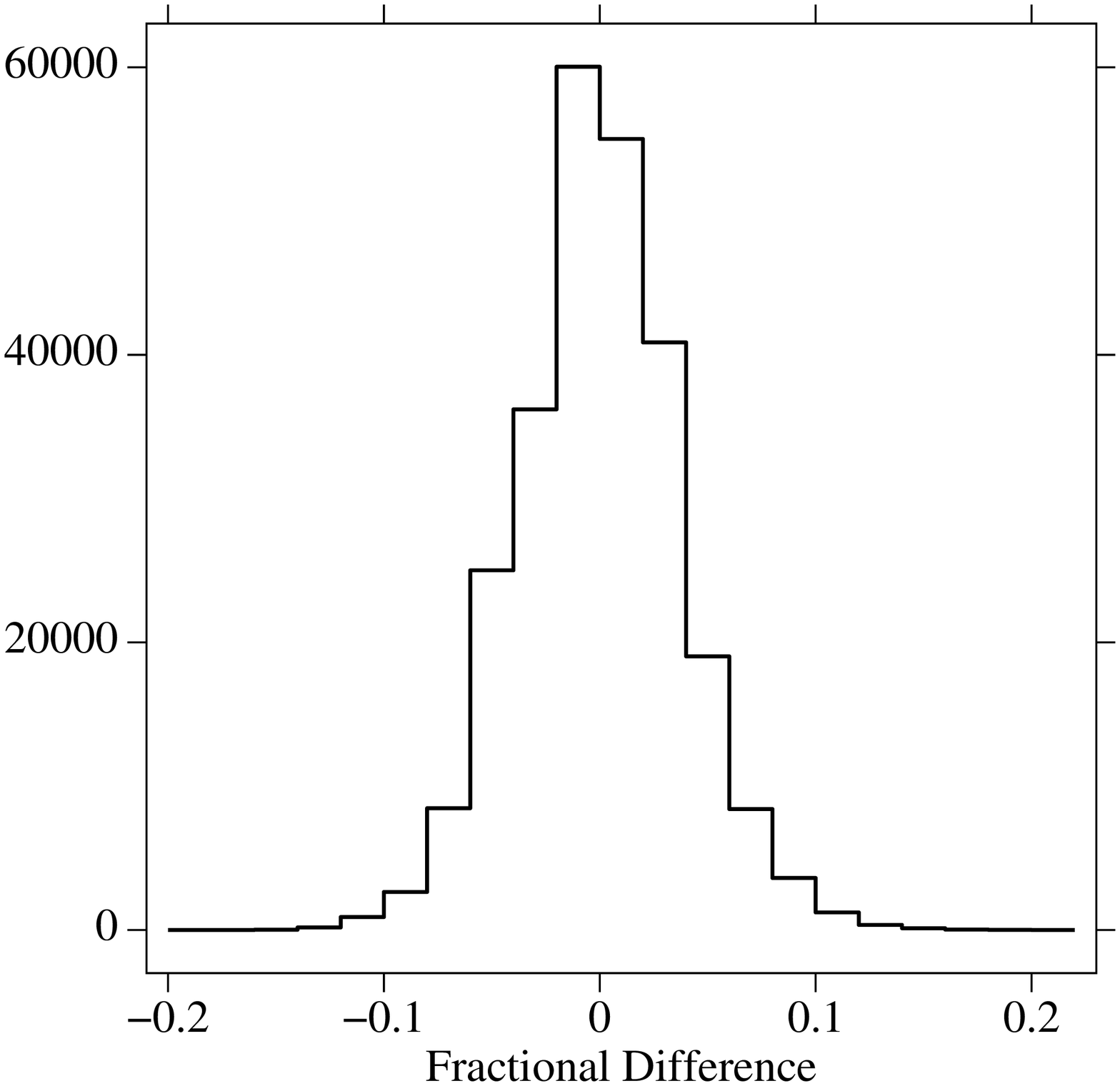} \\
    (c)&(d)\\
    \includegraphics[width=0.46\columnwidth]{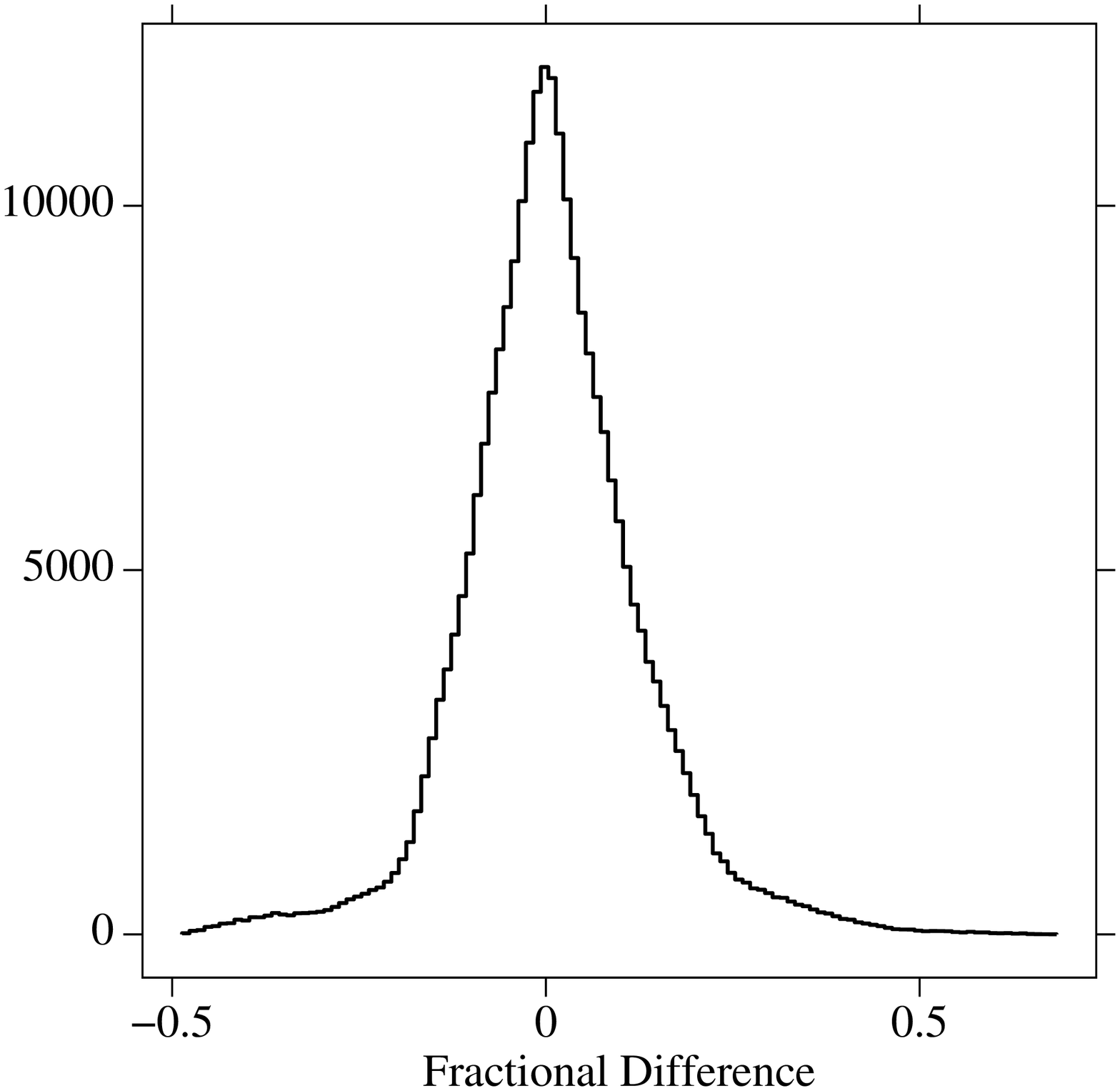} &
    \includegraphics[width=0.46\columnwidth]{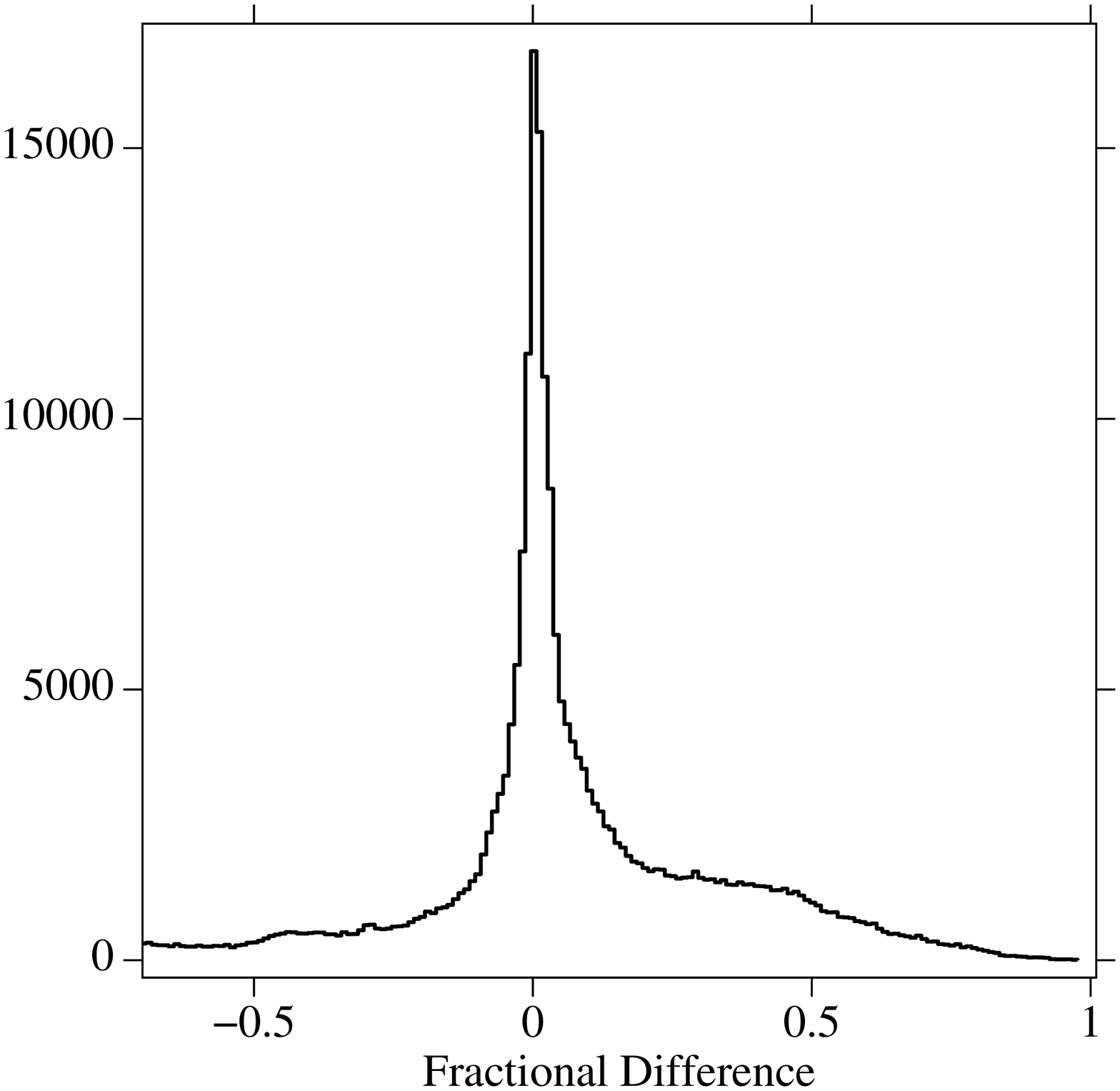} \\
  \end{tabular}

  \caption{The distribution of fractional differences between pixels
    in binned images. (a) is between an adaptively binned (0.02
    fractional error) image and the original surface brightness image
    binned with the same bins. (b) is the same as (a) except using
    0.06 fractional error image. (c) is between an adaptively binned
    image (0.02 fractional error) and the original surface brightness
    image. (d) is between a $\sigma_{\mathrm{min}}=4$ AS image and the
    original surface brightness image.}
  \label{fig:simcluster_hist}
\end{figure}

A more useful and quantitative analysis of the performance of the
adaptive binning algorithm can be made by plotting the fractional
differences (the absolute value of which is in the above images) as
histograms, shown in Fig.  \ref{fig:simcluster_hist}. (a) displays the
difference between the 0.02 error adaptively binned image and the
surface brightness image binned using the same bins \emph{for each
  pixel}. (b) is the same as (a), except using 0.06 error binned
images.  (c) shows a histogram of the differences between the 0.02
error adaptively binned image and the unbinned surface brightness
image. The difference between an image produced by \textsc{asmooth}
(Fig.  \ref{fig:simcluster}(d)) and the surface brightness image is
shown in (d) for comparison.

Fig. \ref{fig:simcluster_hist}(a) and (b) approximate Gaussian
distributions. The widths of the distributions are close to the
threshold fractional error values used by the algorithm, 0.02 and
0.06.  (c) shows a wider distribution, which is obtained due to the
count varying across the bin in the surface brightness image, but not
in the adaptively binned image. The results in (a) and (b) give
confidence in the effectiveness of the algorithm, as the distributions
are symmetric and Gaussian.

The histogram for the AS image in (d) has a very narrow peak, but
there are long tails, mainly due to the edge-effects. Excluding the
outer region almost removes the tail.

We also analysed the differences between the input surface brightness
image and the adaptively binned image using the $\chi^2$ statistic. If
there are $S_i$ total counts in the surface brightness image in bin
$i$, and $s_i$ in the adaptively binned image, and $n_i$ pixels in the
bin, then the statistic is
\begin{equation}
  \chi^2_i = \frac{(S_i-s_i)^2}{S_i+n_i b}.
\end{equation}
We plotted the distribution of $\chi^2$ for the bins in the adaptively
binned image. It showed reasonable agreement with the distribution
predicted for one degree of freedom.

\section{Colour binning}
A simple way to investigate the inner structure of a cluster is to
create X-ray images in different energy bands. The images can then be
divided to show the areas of hard and soft X-ray emission, known as a
colour image.  Additionally plasma codes, such as the \textsc{mekal}
thermal model, can be fitted to the relative counts, enabling one to
map temperature, metallicity and absorption.

One faces a similar problem to the simple intensity binning problem.
How can ratios of counts be formed so that the error on the result is
accurate enough? A binning algorithm must be able to use the same size
bins on the images to be divided, plus it must adapt its bin size to
the counts in each of the bands, rather than just one.

The colour adaptive binning algorithm is similar to the intensity
algorithm, except it uses an input image for each band, to produce an
output image for each band. Each input image is binned using the same
bins. The bins are defined by the error on a `combined colour', which
folds the ratios of all the bands, not the error on the intensity. If
$s^i_k$ is the background-subtracted count in band $i$ for bin $k$,
the combined colour is defined as
\begin{equation}
  C_k =
  \left[ \left( s^1_k/s^2_k \right)/s^3_k \right] / \cdots.
\end{equation}
The fractional error on the combined colour, if there are $N$ bands,
is
\begin{equation}
  \frac{\sigma_{ \left( C_k \right) }}{C_k} =
  \left[
    \sum_{i=1}^{N}{ \frac{ c^i_k + n_k b^i }
      { \left( c^i_k - n_k b^i \right)^2 } }
  \right]^{1/2}.
\end{equation}
This error is a useful quantity because it takes into account the
count in each band in a symmetric way. To make an actual colour map,
two binned output images are divided.

An alternative approach is to adaptively bin a total intensity image.
The resultant bin-map can then be applied individually to each of the
X-ray bands. Ratios can be formed by dividing the output images.  This
approach has its advantages when several bands are being used. If a
sharp feature is only present in one band, then the source will be
binned with a large bin using the combined-colour method, due to the
large statistical uncertainty on the colour. However, binning using
the total intensity leads to colours with larger range of errors than
the combined-colour approach.

\subsection{Real example}
The combined colour adaptive binning method has been applied to the
Perseus data shown earlier. Three images of the cluster were created
in different bands, labelled from 0--2, in the energy ranges 0.5--1,
1--2 and 2--7\keV.

The colour adaptive binning algorithm above was applied to the data
using a fractional errors on the combined colour, $\left( s^0/s^1
\right)/s^2$, of 0.12 and 0.2. Again the background counts in each
image were small enough to be neglected.

\begin{figure}
  \begin{center}
    \includegraphics[width=0.95\columnwidth]{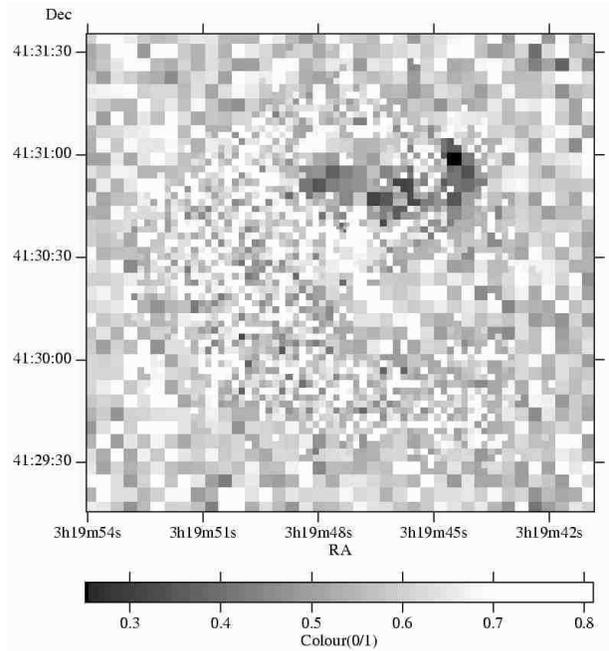}
  \end{center}
  \caption{Colour map showing the ratio of counts in band 0 (0.5--1\keV)
    to band 1 (1--2\keV) for the Perseus cluster, adaptively binned
    with a fractional error of 0.2. This colour map highlights regions
    of high photoelectric absorption.}
  \label{fig:per_col01}
\end{figure}

Fig. \ref{fig:per_col01} shows the $s^0/s^1$ colour image produced
with a fractional error of 0.2, which is sensitive to X-ray
absorption; darker shades indicate more absorption.  The absorption
shadow of an infalling dwarf galaxy, discussed by Fabian et al.
(2000), is clearly seen close to the centre of the image. The rest of
the image has approximately uniform colour, indicating there are no
further strong absorption features.

\begin{figure}
  \begin{center}
    \includegraphics[width=0.95\columnwidth]{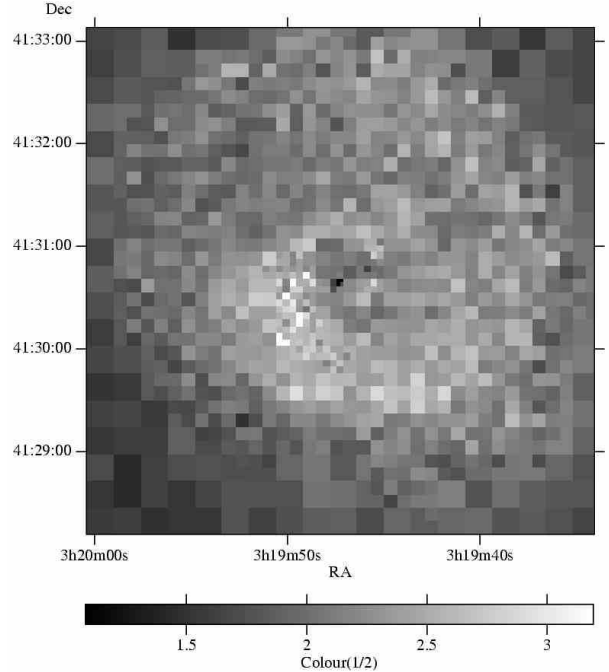}
  \end{center}
  \caption{Colour map showing the ratio of counts in band 1 (1--2\keV)
    to band 2 (2--7\keV) for the Perseus cluster, adaptively binned
    with a fractional error of 0.12. This colour map highlights
    regions of low temperature.}
  \label{fig:per_col12}
\end{figure}

Fig. \ref{fig:per_col12} shows the $s^1/s^2$ colour image produced
with a fractional error of 0.12, which is sensitive to temperature;
lighter shades indicate lower temperature. From this simple analysis,
there appears to be a temperature gradient in the cluster, with cooler
temperatures towards the centre and the brighter regions. This agrees
with cooling flow models. The analysis emphasises a spiral structure,
which is present right down to the cluster centre (compare with Fig.
\ref{fig:per_acisi_ab}).

Using the combined colour method to generate the bins, the fractional
errors on the bins for the individual colours, both range from 0.05 to
0.1 using a fractional error of 0.12. We have shown the colours with
different fractional errors due to the relative strengths of features
in the colours.

\subsection{Spectral fitting}
As mentioned above, one can fit the relative counts in different X-ray
bands against the relative counts predicted by a plasma code, to
estimate physical properties, such as temperature and metallicity. To
fit many parameters, several bands are required. It is useful to
choose bands with approximately the same number of counts, the choice
of which can be derived from the spectrum.

For a more detailed analysis, full spectral fitting is necessary. We
have investigated taking an intensity-binned image of the object,
automatically extracting spectra for each of the bins, automating
\textsc{xspec} to fit each spectrum, and then plotting the fitted
parameters spatially. As a side issue, the data must be binned so that
there are enough counts per pixel to assume Gaussian errors. Also,
spatial variations of the response matrix of the detector and
telescope may need to be considered.

\section{Availability}
An implementation of the adaptive binning algorithm is available
written in the C++ programming language. See the URL
\url{http://www-xray.ast.cam.ac.uk/~jss/adbin/} for instructions for
download.

\section{Summary}
We have presented an algorithm to adaptively bin data. We have
demonstrated its use on X-ray intensity and colour images of the
Perseus cluster. We also applied the algorithm to simulated data. The
method has also been used to create multi-layer colour images, where
the colours represent the intensity in different bands. The fractional
error threshold can be varied depending on how much statistical noise
is acceptable in the analysed image.

The method is not limited to binning on intensity or colour. Any sort
of data may be binned, providing there is a method to compute the
statistical fractional error, or weighting, of a group of pixels. The
algorithm may be useful in presenting the output from numerical
simulations.

The method is also not limited to two-dimensional data. We have used
it already on one-dimensional cuts through images. It may also be
simply extended to work on $N$-dimensional data sets.

\section*{Acknowledgements}
ACF and JSS thank the Royal Society and PPARC for support,
respectively.

\end{document}

%% file: defn.tex
\def\keV{{\rm\thinspace keV}}

\def\Msun{\hbox{$\rm\thinspace M_{\odot}$}}

\def\yr{{\rm\thinspace yr}}

\def\Msunpyr{\hbox{$\Msun\yr^{-1}\,$}}